\documentclass[UTF8,11pt]{article}
\usepackage{amsmath}
\usepackage{graphicx}
\usepackage{float}
\usepackage{graphicx,subfigure}

\usepackage{geometry}
\usepackage{a4wide}
\usepackage{amsmath}
\usepackage[
      colorlinks=true,
      linkcolor=blue,
      urlcolor=blue,
      filecolor=black,
      citecolor=blue,
            ]{hyperref}
\usepackage{color}
\usepackage{amsfonts}
\usepackage{amssymb}
\usepackage{epstopdf}
\usepackage{caption}
\usepackage{subfigure}
\usepackage{multirow}
\usepackage{makecell}
\usepackage{cite}
\usepackage{nicefrac}

\makeatletter
\newcommand\figcaption{\def\@captype{figure}\caption}
\newcommand\tabcaption{\def\@captype{table}\caption}
\makeatother
\begin{document}

\title{\Large \textbf{Inside Anisotropic Black Hole with Vector Hair}}
\author{\large
Rong-Gen Cai$^{a,b,c}$\footnote{cairg@itp.ac.cn}~,
~~Chenghu Ge$^{\,a,b}$\footnote{gechenghu@itp.ac.cn}~,
~~Li Li$^{\,a,b,c}$\footnote{liliphy@itp.ac.cn (corresponding author)}~,
~~Run-Qiu Yang$^{d}$\footnote{aqiu@tju.edu.cn (corresponding author)}
\\
\\
\small $^a$CAS Key Laboratory of Theoretical Physics, Institute of Theoretical Physics, \\
\small Chinese Academy of Sciences, Beijing 100190, China\\
\small $^b$School of Physical Sciences, University of Chinese Academy of Sciences, \\
\small No.19A Yuquan Road, Beijing 100049, China\\
\small $^c$School of Fundamental Physics and Mathematical Sciences, Hangzhou Institute for Advanced Study, \\
\small UCAS, Hangzhou 310024, China\\
\small $^d$Center for Joint Quantum Studies and Department of Physics, School of Science, Tianjin University, \\
\small Yaguan Road 135, Jinnan District, 300350 Tianjin, China
}
\date{}

\maketitle

\begin{abstract}
We study the internal structure of anisotropic black holes with charged vector hairs. Taking advantage of the scaling symmetries of the system, some radially conserved charges are found via the extension of the Noether theorem.  Then, a general proof of no inner horizon of these black holes is presented and the geometry ends at a spacelike singularity. Before reaching the singularity, we find several intermediate regimes both analytically and numerically. In addition to the Einstein-Rosen bridge contracting towards the singularity, the instability triggered by the vector hair results in the oscillations of vector condensate and the anisotropy of spatial geometry. Moreover, the latter oscillates at twice the frequency of the condensate. Then, the geometry enters into Kasner epochs with spatial anisotropy. Due to the effects from vector condensate and U(1) gauge potential,  there is generically a never-ending alternation of Kasner epochs towards the singularity. The character of evolution on approaching the singularity is found to be described by the Kasner epoch alternation with flipping of powers of the
Belinskii-Khalatnikov-Lifshitz type.
\end{abstract}

\newpage
\tableofcontents

\section{Introduction}
Strong observational evidence for the existence of black holes predicted by Einstein's general relativity has been accumulating~\cite{LIGOScientific:2016aoc,EventHorizonTelescope:2019dse,EventHorizonTelescope:2019ths}, which raises many conceptually important questions that beg for answers and motivate future research.
Of particular interest is the black hole internal structure for which many of the deep conceptual issues appear. Since the singularity theorem proved by Penrose~\cite{Penrose:1964wq}, it was widely accepted that in the interior of a black hole there typically lies a singularity where the spacetime curvature diverges in most cases. An outstanding result was obtained by Belinskii, Khalatnikov and Lifshitz (BKL) for the generality of singularities~\cite{Lifshitz:1963ps,Belinsky:1970ew,Belinski:1973zz}: as approaching the singularity, there is a generic chaotic behavior described by an infinite alternation of the so-called Kasner epochs. It is worth emphasizing that from the BKL observation it does not yet follow that the chaotic singularity is inevitable. While BKL showed that the initial data for BKL singularity represent a set of nonzero measures in the space of all possible data, there can be many general solutions of different types, including a general solution avoiding oscillating behavior near the singularity, see \emph{e.g.}~\cite{Belinski:1973zz,Andersson:2000cv,Erickson:2003zm,Ivashchuk:2005kn}.

Apart from the singularity, another important question is the number of horizons inside a black hole. Some analytical black hole solutions with different number of horizons were obtained in the literature. For example, the Schwarzschild black hole has one horizon and the Kerr-Newman black hole has at most two horizons. One is also able to construct a black hole with arbitrary number of horizons~\cite{Nojiri:2017kex,Gao:2021kvr}. The inner horizon behind the event horizon is known as a Cauchy horizon at which the predictability of the classical dynamics of general relativity breaks down\,\footnote{The Cauchy horizon is defined as the boundary of the domain of dependence of a maximal spacelike hypersurface outside the black hole.}, even far away from any singularity. To keep the predictability of the classical dynamics, the strong cosmic censorship (SCC) conjecture asks for the instability and ensuing disappearance of the Cauchy horizon from generic initial data~\cite{Ringstrom:2015jza,Isenberg:2015rqa}. While the BKL approach focuses on asymptotic behavior in the vicinity to the singularity and the SCC has to do with the global structure of spacetime, both play significant roles in our understanding of the internal structure of black holes.~\footnote{It should be stressed here that the mathematically rigorous formulation of the SCC conjecture and the proof of BKL conjecture are still open problems.}

In addition to the smoking-gun observational evidence of black holes, another motivation comes from holography which provides the reinterpretation of  generic gravitational phenomena in terms of field theory in the boundary of spacetime, and vice versa. The extra dimension in the gravity description corresponds to the energy scale of the dual field theory, and thus a gravitational solution provides a geometric realization of renormalization group (RG) flow of the field theory. This flow can go smoothly through the event horizon towards singularity. In the absence of an inner Cauchy horizon as supported by the SCC conjecture, the holographic flow will approach a cosmological singularity that emerges dynamically in the black hole interior. While to what extent can holography shed light on the nature of spacelike singularities is still under investigation, the holography has provided some useful probes for the black hole interior, such as correlation functions~\cite{Fidkowski:2003nf,Festuccia:2005pi}, entanglement entropy~\cite{Hartman:2013qma} and complexity~\cite{Stanford:2014jda,Brown:2015bva}.

Recently, some interesting dynamics inside black hole with charged scalar hairs have been classified. On the one hand, it has been generally proved that there is no Cauchy horizon for those hairy black holes~\cite{Cai:2020wrp,An:2021plu}. On the other hand, for the asymptotically anti-de Sitter (AdS) solutions known as holographic s-wave superconductors, the interior evolves through several distinct epochs~\cite{Hartnoll:2020fhc}, including a collapse of the Einstein-Rosen (ER) bridge, Josephson oscillations of the scalar field and the final Kasner singularity. At a discrete set of temperatures, an infinite number of Kasner inversions was observed numerically. In contrast to the BKL analysis, such Kasner inversions are induced by the charge of the scalar field and thus are sensitive to the black hole temperature. While the existence of the final Kasner universe has been shown to be sensitive to the potential of the scalar field~\cite{Cai:2020wrp,An:2021plu}, other epochs seem to be general features closely associated with the instability of the inner horizon of charged black holes without scalar hair. See other studies of black holes for which the Cauchy horizon is removed by scalar hairs~\cite{Hartnoll:2020rwq,Devecioglu:2021xug,VandeMoortel:2021gsp,Mansoori:2021wxf,Grandi:2021ajl}. Thus far, the structure as well as dynamical epochs inside the event horizon was studied by introducing a scalar field to black holes with maximally symmetric horizon.\footnote{See recent generalization of the constraint on the number of horizons to the static but inhomogeneous case by considering energy conditions~\cite{Yang:2021civ}, and to a certain class of stationary black holes in which the inner horizon cannot exist~\cite{Dias:2021afz}.} Nevertheless, it is remarkable that, even in such a simple setup, the onset of a scalar hair outside the event horizon is closely related to significantly intricate dynamics inside. It also suggests that the Cauchy horizon in large classes of theories will be definitely removed by a non-trivial scalar hair, thus the SCC is respected and the geometry approaches a cosmological singularity that would be BKL-like or beyond.

In this work, we will consider the interior of a large class of planar black holes with a charged vector field. The vector hair can be generated spontaneously, resulting an anisotropic black hole with non-trivial vector hairs. It provides a holographic realization of p-wave superconductivity~\cite{Cai:2013pda,Cai:2013aca}. Another p-wave model was constructed by considering the SU(2) Yang-Mills field~\cite{Gubser:2008wv}. We will prove that the latter is a special case of the former charged vector theory. Comparing with the scalar case~\cite{Cai:2020wrp,An:2021plu,Hartnoll:2020fhc}, one needs to turn on one more metric component to account for the breaking of rotation symmetry due to the appearance of the vector hair. Thus it is interesting to see whether the Cauchy horizon exists or not for the vector case and the dynamics inside those hairy black holes. Remarkably, we are able to prove that there is no inner Cauchy horizon for the black holes with non-trivial vector hair. This is valid for models with quite general couplings, including, in particular, the Gauss-bonnet term.  We will also uncover novel nonlinear dynamics inside the hairy black holes. In contrast to the scalar case, without any parameter fine tuning, there is a never-ending oscillating behavior of the BKL type near the spacelike singularity, which will be shown to be the contribution of either vector condensate or U(1) gauge potential.

The rest of the paper is organized as follows. In Section~\ref{sec:setup}, we introduce the charged vector model and construct the Noether charge associated with the scaling symmetries of the system. A general proof of no inner horizon of black hole with vector hair is presented in Section~\ref{sec:proof}. In Section~\ref{sec:interior}, we discuss in detail the dynamic epochs inside the hairy black hole, for which some new features will be uncovered compared to the scalar case. We conclude with some discussion in Section~\ref{sec:discussion}. In Appendix~\ref{app:YMs}, we show that the Einstein-SU(2) Yang-Mills model is a special case of the charged vector model we consider.  Appendix~\ref{app:charge} is devoted to the Noether charge for system with higher derivatives. In appendix~\ref{app:At}, we prove that the gauge potential should vanish at every horizon of the hairy black holes.

\section{Model and Noether Charge}\label{sec:setup}
We introduce a charged vector field $\rho_\mu$ into the $(d+2)$ dimensional Einstein-Maxwell theory with a negative cosmological constant $\Lambda$. We begin with the  action~\cite{Cai:2013pda,Cai:2013aca}
\begin{equation}\label{action}
\begin{split}
S=\frac{1}{2\kappa_N^2}\int d^{d+2} x
\sqrt{-g}(\mathcal{R}-2\Lambda+\mathcal{L}_m),\\
\mathcal{L}_m=-\frac{1}{4}F_{\mu\nu} F^{\mu \nu}-\frac{1}{2}\rho_{\mu\nu}^\dagger\rho^{\mu\nu}-m^2\rho_\mu^\dagger\rho^\mu+iq\gamma \rho_\mu\rho_\nu^\dagger F^{\mu\nu},
\end{split}
\end{equation}
where $F_{\mu\nu}=\nabla_\mu A_\nu-\nabla_\nu A_\mu$ and $\rho_{\mu\nu}=D_\mu\rho_\nu-D_\nu\rho_\mu$ with $D_\mu=\nabla_\mu-iq A_\mu$. The last non-minimal coupling term characterizes the magnetic moment of the vector field $\rho_\mu$, which plays an important role in the case with an applied magnetic field~\cite{Cai:2013pda,Cai:2013kaa}. 

Varying the action, we obtain the equations of motion for matter
\begin{eqnarray}
\label{gauge}
&& \nabla^\nu F_{\nu\mu}=iq(\rho^\nu\rho_{\nu\mu}^\dagger-{\rho^\nu}^\dagger\rho_{\nu\mu})+iq\gamma\nabla^\nu(\rho_\nu\rho_\mu^\dagger-\rho_\nu^\dagger\rho_\mu),
\\
\label{vector}
&& D^\nu\rho_{\nu\mu}-m^2\rho_\mu+iq\gamma\rho^\nu F_{\nu\mu}=0,
\end{eqnarray}
as well as for metric
\begin{equation}\label{tensor}
\begin{split}
\mathcal{R}_{\mu\nu}-\frac{1}{2}\mathcal{R}g_{\mu\nu}=T_{\mu\nu},
\end{split}
\end{equation}
with the energy momentum tensor
\begin{equation}\label{energy momentum tensor}
	\begin{split}
		T_{\mu\nu}=&-\Lambda g_{\mu\nu}+\frac{1}{2}F_{\mu\lambda}{F_\nu}^\lambda+\frac{1}{2}\mathcal{L}_m g_{\mu\nu}\\
		&+\frac{1}{2}\{[\rho_{\mu\lambda}^\dagger{\rho_\nu}^\lambda+m^2\rho_\mu^\dagger\rho_\nu-iq\gamma(\rho_\mu\rho_\lambda^\dagger-\rho_\mu^\dagger\rho_\lambda){F_\nu}^\lambda]+\mu\leftrightarrow\nu\}.
	\end{split}
\end{equation}
Here we have absorbed the cosmological constant term to $T_{\mu\nu}$.

We are going to find planar charged black hole solutions to the theory~\eqref{action} by taking the following ansatz:
\begin{equation}\label{ansatz}
\begin{split}
ds^2=\frac{1}{z^2}\left[-f(z)e^{-\chi(z)}dt^2+\frac{dz^2}{f(z)}+u(z)dx^2+d\vec{y}_{d-1}^2\right]\,,\\
\rho_\nu dx^\nu=\rho_x(z)dx,\quad A_\nu dx^\nu=A_t(z)dt\,.
\end{split}
\end{equation}
Here the AdS boundary is at $z=0$ and the singularity will be at $z\rightarrow\infty$. The gauge symmetry and equation of motion~\eqref{vector} ensure that the phase factor of $\rho_x$ is constant, so we can take $\rho_x$ to be real everywhere. The temperature of the black hole is given by
\begin{equation}
T=-\frac{f'(z_H)e^{-\chi(z_H)/2}}{4\pi}\,,
\end{equation}
where $z_H$ is the location of the event horizon at which $f(z_H)=0$.

The independent equations of motion now read
\begin{equation}\label{eoms}
\begin{split}
\left(\frac{e^{\chi/2}\sqrt{u}A_t'}{z^{d-2}}\right)'=\frac{2q^2e^{\chi/2}\rho_x^2A_t}{z^{d-2}f\sqrt{u}}\,,\\
\left(\frac{e^{-\chi/2}f\rho_x'}{z^{d-2}\sqrt{u}}\right)'=-\frac{q^2e^{\chi/2}\rho_xA_t^2}{z^{d-2}f\sqrt{u}}+\frac{m^2e^{-\chi/2}\rho_x}{z^{d}\sqrt{u}}\,,\\
u''+\left(\frac{f'}{f}-\frac{u'}{2u}-\frac{\chi'}{2}-\frac{d}{z}\right)u'=-2z^2\rho_x'^2-\frac{2m^2\rho_x^2}{f}+\frac{2q^2z^2e^{\chi}\rho_x^2A_t^2}{f^2}\,,\\
\frac{d}{2}\chi'-d\frac{f'}{f}-\frac{d}{2}\frac{u'}{u}+\frac{d(d+1)}{z}=\frac{z^3\rho_x'^2}{u}-\frac{2\Lambda}{zf}-\frac{z^3e^{\chi} A_t'^2}{2f}-\frac{q^2z^3e^{\chi}\rho_x^2A_t^2}{f^2u}\,,\\
\left(\frac{d}{z}+\frac{u'}{2u}\right)\frac{f'}{f}-\frac{u'}{2u}\chi'-\frac{d(d+1)}{z^2}=-\frac{z^2\rho_x'^2}{u}+\frac{2\Lambda}{z^2f}-\frac{m^2\rho_x^2}{fu}+\frac{z^2e^{\chi}A_t'^2}{2f}+\frac{3q^2z^2e^{\chi}\rho_x^2A_t^2}{f^2u}\,,
\end{split}
\end{equation}
where the prime denotes the derivative with respect to the radial coordinate $z$. Note that the geometry is homogeneous but anisotropic due to the formation of a vector field along the $x$-direction. We prove in Appendix~\ref{app:YMs} that the Einstein-SU(2) Yang-Mills theory~\cite{Gubser:2008wv} is a special case of the Einstein-Maxwell-vector theory and yields the same equations of motion~\eqref{eoms} with $m^2=0$.

It is obvious that a charged black hole without vector hair is described by the AdS-Reissner-Nordstr\"{o}m (RN) black hole
\begin{equation}\label{RN}
\begin{split}
ds^2&=\frac{1}{z^2}\left[-f(z)dt^2+\frac{dz^2}{f(z)}+dx^2+d\vec{y}_{d-1}^2\right], \qquad\quad\;\, A_t(z)=\mu\left[1-\left(\frac{z}{z_H}\right)^{d-1}\right]\,,\\
f(z)&=-\frac{2\Lambda}{d(d+1)}\left[1-\left(\frac{z}{z_H}\right)^{d+1}\right]-\frac{(d-1) \mu ^2 {z_H}^2}{2 d}\left(\frac{z}{z_H}\right)^{d+1}\left[1-\left(\frac{z}{z_H}\right)^{d-1}\right]\,,
\end{split}
\end{equation}
where $z_H$ and $\mu$ represent, respectively, the event horizon and the chemical potential. The solution has an inner Cauchy horizon. An interesting question arises: will the inner horizon be removed by non-trivial vector hairs? Since the resulting non-linear coupled equations do not have analytical solutions, one may try to solve the system numerically and check whether the inner horizon exists or not. This is complicated in numerics and is only able to check some particular cases. Instead of numerically solving the coupled equations~\eqref{eoms}, one urges a general proof. Inspired by our previous work~\cite{Cai:2020wrp,An:2021plu}, we wonder if it is possible to construct a radially conserved quantity for the anisotropic geometry of~\eqref{ansatz}. In the remaining part of this section, we will show how to obtain the conserved quantity by taking advantage of the scaling symmetry of the system.

Substituting the ansatz~\eqref{ansatz} into the action~\eqref{action}, one obtains the following effective action:
\begin{equation}\label{actioneff}
\begin{split}
S_{\text{eff}}&=\frac{\mathcal{V}_{d+1}}{2\kappa_N^2}\int dz \mathcal{L}_{\text{eff}}(f,f',f''; \chi,\chi',\chi''; u,u',u''; A_t, A_t'; \rho_x, \rho_x'; z)\,,\\
\mathcal{L}_{\text{eff}}&=-\frac{ e^{-\chi/2}\sqrt{u}}{z^{d+2}}\left[(d+1) (d+2) f+2 \Lambda \right]+\frac{(d+1) e^{-\chi/2}\sqrt{u}}{ z^{d+1} }\left(2f'-f\chi'+f\frac{u'}{u}\right)\\
&\quad+\frac{f e^{-\chi/2}\sqrt{u}}{z^{d}}\left(\frac{3\chi'}{2}\frac{f'}{f}-\frac{f'}{f}\frac{u'}{u}+\frac{\chi'}{2}\frac{u'}{u}-\frac{\chi'^2}{2}+\frac{u'^2}{2u^2}-\frac{f''}{f}+\chi''-\frac{u''}{u}\right)\\
&\quad+\frac{q^2 e^{\chi/2} \rho_x^2A_t^2}{ z^{d-2}f\sqrt{u}}-\frac{m^2e^{-\chi/2}\rho_x^2}{z^d\sqrt{u}}+\frac{ e^{\chi/2}\sqrt{u} A_t'^2}{2z^{d-2} }-\frac{ f e^{-\chi/2} \rho_x'^2}{z^{d-2}\sqrt{u}}\,,\\
\end{split}
\end{equation}
with $\mathcal{V}_{d+1}=\int dtdxd^{d-1}y$. It is clear that $\mathcal{L}_{\text{eff}}$ depends on five real functions $\mathcal{F}=\{f, \chi, u, A_t,\rho_x\}$, of variable $z$, and on the derivatives of $\{f, \chi, u\}$ up to second order. Moreover, $\mathcal{L}_{\text{eff}}$ depends explicitly on $z$.

The key observation is that there is a scaling symmetry associated with $S_{\text{eff}}$ as follows,
\begin{equation}\label{scaling1}
\begin{split}
z\rightarrow \lambda z,\quad f(z)\rightarrow f(z),\quad\chi(z)\rightarrow \chi(z)-2d\log(\lambda),\quad u(z)\rightarrow \lambda^2 u(z),\\
A_t(z)\rightarrow \lambda^{d-1} A_t(z),\quad \rho_x(z)\rightarrow\rho_x(z)\,,
\end{split}
\end{equation}
where $\lambda$ is a positive constant. We then construct the corresponding Noether charge for $S_{\text{eff}}$ which contains higher derivatives. Let's consider an infinitesimal transformation for the continuous symmetry~\eqref{scaling1}, \emph{i.e.} $\lambda=1+\epsilon$ with $\epsilon$ a small parameter. The variations of radial coordinate and the  functions that are first-order small are given by
\begin{equation}
\delta z=\epsilon Z(z), \quad \delta \mathcal{F}_a(z)=\epsilon F_a(z)\,,
\end{equation}
with
\begin{equation}\label{scaling2}
\begin{split}
Z=z, \quad F_f=0, \quad F_{\chi}=-2d,\quad F_u=2u(z),\quad F_{A_t}=(d-1)A_t(z), \quad F_{\rho_x}=0\,.
\end{split}
\end{equation}
A straightforward extension of the proof of the Noether theorem now yields a radially conserved Noether charge~\footnote{See Appendix~\ref{app:charge} for a general proof of the conserved Noether charge used in~\eqref{initial charge}.}
\begin{equation}\label{initial charge}
\begin{split}
\mathcal{Q}_{\text{Noether}}&=Z\mathcal{L}_{\text{eff}}+\sum_{a=\{f, \chi, u, A_t,\rho_x\}} \left[\frac{\partial\mathcal{L}_{\text{eff}}}{\partial\mathcal{F}_a'}-\frac{d}{dz}\left(\frac{\partial\mathcal{L}_{\text{eff}}}{\partial\mathcal{F}_a''}\right)\right](F_a-Z\mathcal{F}_a')+\left(\frac{\partial\mathcal{L}_{\text{eff}}}{\partial\mathcal{F}_a''}\right)\frac{d(F_a-Z\mathcal{F}_a')}{dz}\,,\\
&=-\frac{d(d+1)fe^{-\chi/2}\sqrt{u}}{z^{d+1}}+\frac{df e^{-\chi/2} \sqrt{u}}{z^d}\left(\frac{f'}{f}-\chi'+\frac{u'}{u}\right)+\frac{ fe^{-\chi/2}\sqrt{u}}{2z^{d-1}}\left(-\frac{f'}{f}\frac{u'}{u}+\chi'\frac{u'}{u}\right)\\
&\quad-\frac{2e^{-\chi/2}\sqrt{u}}{z^{d+1}} \Lambda+\frac{q^2e^{\chi/2} \rho_x^2 A_t^2}{ z^{d-3} f\sqrt{u}}-\frac{m^2 e^{-\chi/2}\rho_x^2}{z^{d-1}\sqrt{u}}-\frac{e^{\chi/2}\sqrt{u} A_t'^2}{2 z^{d-3}}+\frac{f e^{-\chi/2}\rho_x'^2}{z^{d-3}\sqrt{u}}\\
&\quad+\frac{e^{\chi/2}\sqrt{u}}{z^{d}}\left[\left(fe^{-\chi}\right)'-z^2A_tA_t'\right](1-d)\,,
\end{split}
\end{equation}
for which $\mathcal{Q}'_{\text{Noether}}(z)=0$. After substituting the $zz$-component of  Einstein's equation~\eqref{tensor}, we obtain a simple expression of the conserved charge:
\begin{equation}\label{conserved charge}
\begin{split}
\mathcal{Q}(z)=\frac{e^{\chi/2}\sqrt{u}}{z^{d}}\left[\left(fe^{-\chi}\right)'-z^2A_tA_t'\right]\,.
\end{split}
\end{equation}
As a double check, we have confirmed that $\mathcal{Q}'=0$ by directly substituting the equations of motion. Note that the conserved charge now depends on the anisotropic factor $u$ of the metric.

Before proceeding to the proof of no inner horizon, we find that the system also allows the following scaling symmetry:
\begin{equation}\label{scaling3}
z\rightarrow \lambda^{-1} z, \quad\chi(z)\rightarrow \chi(z)+2\log(\lambda),\quad u(z)\rightarrow \lambda^{-2d} u(z),\quad \rho_x(z)\rightarrow\lambda^{1-d}\rho_x(z)\,,
\end{equation}
with $f$ and $A_t$ unchanged. Similarly, we can obtain another radially conserved quantity
\begin{equation}\label{charge2}
\widetilde{\mathcal{Q}}(z)=\frac{fe^{-\chi/2}}{z^{d}\sqrt{u}}\left(u'+2z^2\rho_x\rho_x'\right)\,.
\end{equation}
%

\section{Proof of no Cauchy Horizon}\label{sec:proof}
In this section we prove the no Cauchy horizon theorem for the black holes with charged vector hairs. Following a similar spirit as in our previous work~\cite{Cai:2020wrp}, we begin with supposing that there is an inner Cauchy horizon at $z_I$ behind the event horizon located at $z_H$. For our present coordinate system~\eqref{ansatz}, we have
\begin{equation}\label{hhorizon}
f(z_H)=f(z_I)=0,\quad z_H<z_I\,.
\end{equation}
For the black hole with non-trivial charged vector hair $\rho_x$, to have a regular horizon, both metric and matter fields are sufficiently smooth near the horizon. The equations of motion~\eqref{eoms} then imply that~\footnote{One immediately obtains that smoothness yields either $A_t=0$ or $\rho_x=0$ at both horizons from the equations of motion. The latter can be shown to be inconsistent with the black hole~\eqref{ansatz} with non-trivial hair $\rho_x$. Refer to Appendix~\ref{app:At} for more details.}
\begin{equation}\label{Athorizon}
A_t(z_H)=A_t(z_I)=0\,,
\end{equation}
with $\rho_x, \chi$ and $u$ finite at both horizons.
\begin{figure}[h!]
\begin{center}
\includegraphics[width=0.7\textwidth]{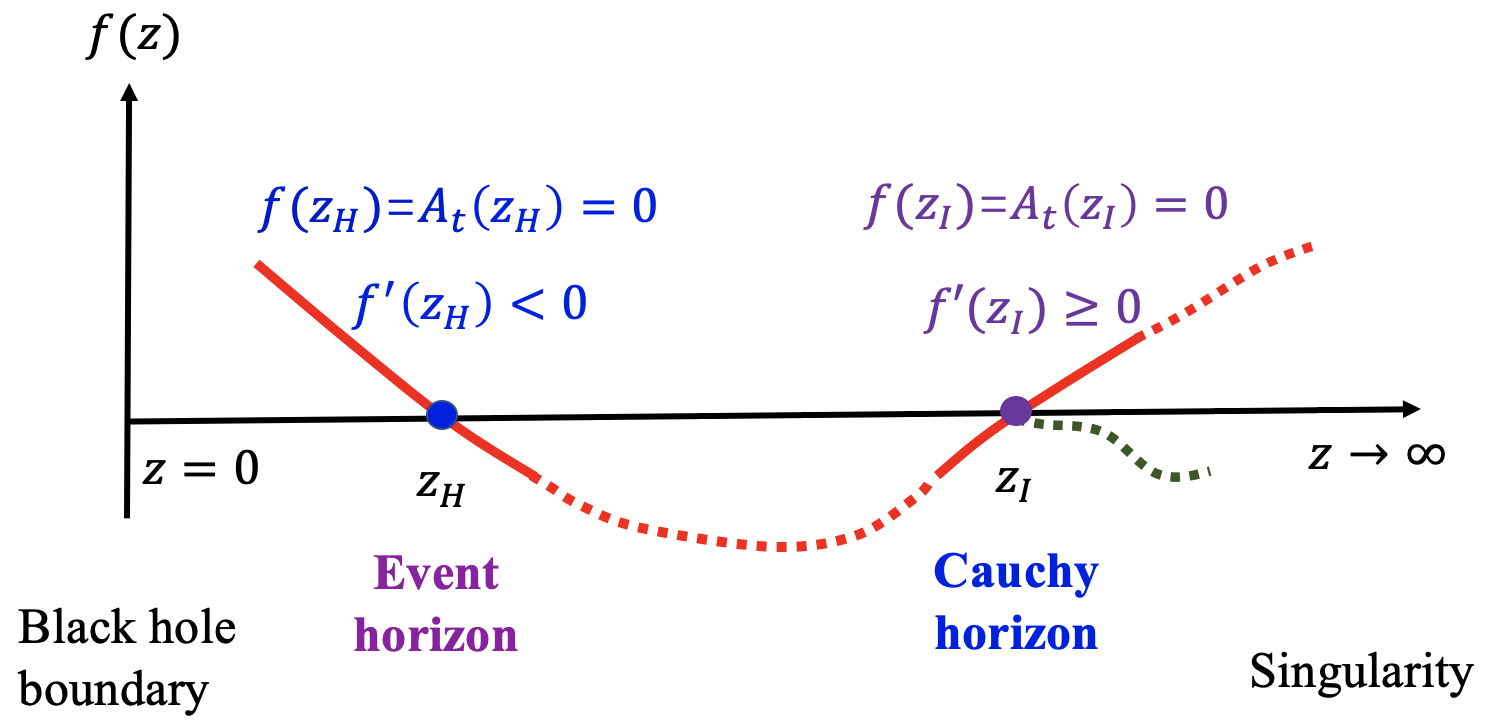}
\caption{Schematic structure of a charged black hole with an inner Cauchy horizon behind the event horizon. The blackening function $f$ and the gauge potential $A_t$ must vanish at the event horizon $z_H$ and the Cauchy horizon $z_I$.}
\label{fig:BH}
\end{center}
\end{figure}

We are interested in the case of black holes with finite temperature \emph{i.e.} $f'(z_H)<0$, for which the structure of the black hole is shown schematically in Figure.~\ref{fig:BH}. The blackening function $f(z)$ goes from positive to negative towards the interior near $z_H$. On the other hand, $f(z)$ turns from negative to positive near the Cauchy horizon $z_I$ (red curve) or $z_I$ is a local maximum (green curve).  Therefore, we obtain
\begin{equation}\label{consth}
f'(z_H)<0,\quad f'(z_I)\ge0\,.
\end{equation}
Computing $\mathcal{Q}$ at both event and inner horizons and using~\eqref{hhorizon} and~\eqref{Athorizon}, we obtain that
\begin{equation}\label{Qhorizon}
\frac{e^{-\chi(z_H)/2}\sqrt{u(z_H)}}{{z_H}^{d}}f'(z_H)=\frac{e^{-\chi(z_I)/2}\sqrt{u(z_I)}}{{z_I}^{d}}f'(z_I)\,.
\end{equation}
From~\eqref{consth} it is clear that the left hand side of~\eqref{Qhorizon} is negative, while the right hand side is non-negative. Therefore, we arrive at a strong result: a smooth inner Cauchy horizon is never able to form for the anisotropic black holes with charged vector hairs. For an arbitrarily small amount of vector hair, the Cauchy horizon ceases to exist and the spacetime ends at a spacelike singularity. Note that~\eqref{Qhorizon} does not depend on $A_t$. Therefore, the present proof is also applicable to the Einstein-vector theory in the absence of Maxwell field $A_\mu$.

\section{Dynamical Epochs Inside the Hairy Black Hole}\label{sec:interior}
After knowing the inner structure behind the event horizon, we are now interested in the dynamics from the horizon to the spacelike singularity. While some dynamical epochs are similar to the charged scalar case~\cite{Hartnoll:2020fhc}, novel features do appear. Just below the critical temperature $T_c$, the vector field will trigger a rapid decrease in $g_{tt}$ towards interior, which is the collapse of the ER bridge.\,\footnote{Note that $t$ becomes a spatial coordinate inside the black hole, thus $g_{tt}$ measures the physical distance for the wormhole connecting the two exteriors of the black hole, \emph{i.e.} the Einstein-Rosen bridge.} There will be a strong oscillation in the vector condensate for which we also call Josephson oscillation as the scalar case. Meanwhile, the function $u(z)$ which characterizes the anisotropy of the geometry will also oscillate but with twice the frequency of the condensate oscillation. After the Josephson oscillation, the solution will enter an era described by anisotropic Kasner universe.  We find that there is a never-ending alternation of Kasner epochs towards the singularity. In contrast to the charged scalar case where the number of transitions from one Kasner epoch to another is very sensitive to black hole parameters, the never-ending oscillating behavior is generic in the vector case we are considering.

To obtain a deep understanding of the interior dynamics, we try to use an analytic approach in addition to numerics. In particular, we will be able to obtain approximate solutions analytically for the epoch of interest by dropping terms that are shown to be negligible from our numerical solutions. The analytic solutions will be compared to the full numerical solutions. We will focus on two features different from the scalar case, including the Josephson oscillations associated with an oscillation of geometry and the Kasner epochs with never-ending alternations.

To illustrate the inner epochs, we consider the p-wave superconductor~\cite{Cai:2013aca} for which the vector condensate $\rho_x$  develops spontaneously. In particular, near the AdS boundary $z=0$ one has
\begin{equation}
A_t=\mu+\dots,\qquad \rho_x=\rho_{x_+}z^{\Delta-1}+\dots,\qquad  \Delta=\frac{d+1+\sqrt{(d-1)^2+4m^2}}{2} \,,
\end{equation}
where $\mu$ and $\rho_{x_+}$ are, respectively, regarded as the chemical potential and vacuum expectation value $\langle \hat{J}^x\rangle$ of the vector operator in the boundary field theory.\,\footnote{In general, the falloff of $\rho_x$ near the AdS boundary behaves as $\rho_x=\rho_{x-} z^{d-\Delta}+ \dots +\rho_{x_+}z^{\Delta-1}+\dots$ with the leading coefficient $\rho_{x-}$ the source of the vector operator $\langle \hat{J}^x\rangle$.} Numerical details for solving the coupled equations of motion can be found in~\cite{Cai:2013aca}. As shown in Appendix~\ref{app:YMs}, under the ansatz~\eqref{ansatz}, the equations of motion for the SU(2) model reduce to the charged vector case~\eqref{eoms} with $m^2=0$. We shall focus on $m^2=0$ and take $\Lambda=-\frac{d(d+1)}{2}$ in our numerics below.

\begin{figure}[htp]
	\begin{center}
		\includegraphics[width=0.55\textwidth]{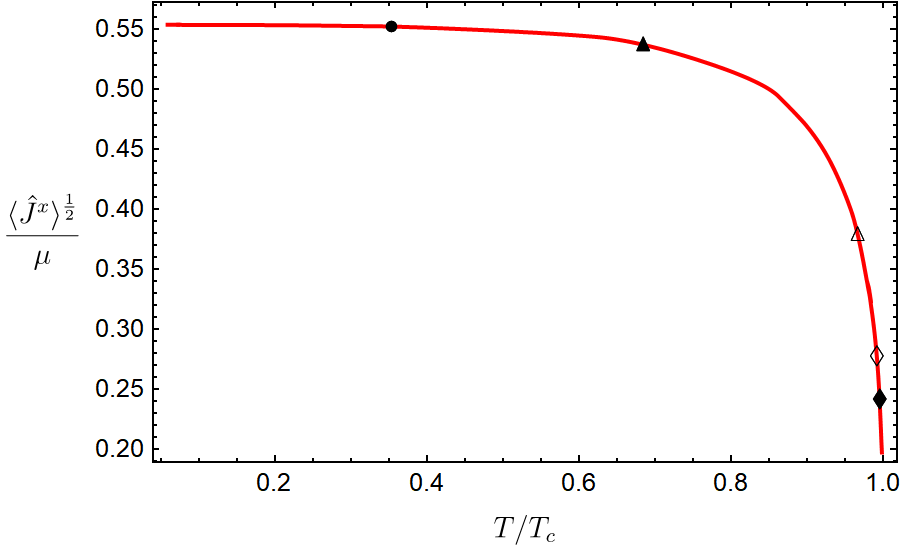}
		\caption{The vector condensate $\langle \hat{J}^x\rangle$ as a function of temperature for holographic p-wave superconductor. Below $T_c$, $\langle \hat{J}^x\rangle$ appears via a second order phase transition. We choose $d=2$, $m^2=0$ and $q=3/2$, and work in units of $\mu$.}
		\label{T-J}
	\end{center}
\end{figure}

The condensate $\langle \hat{J}^x\rangle$ as a function of temperature is shown in Figure.~\ref{T-J} for the four dimensional spacetime $(d=2)$ and $q=3/2$. It is clear that the condensate $\langle \hat{J}^x\rangle$ develops smoothly below a critical temperature $T_c$ and increases monotonically as the temperature is decreased, corresponding to a second order phase transition from the normal phase (RN black hole) to the broken phase (hairy black hole), known as holographic p-wave superconductor~\cite{Cai:2013pda,Cai:2013aca}. In the following discussion, we will construct the holographic flows triggered by the vector condensate and analyze, in particular, the asymptotic behavior near the spacelike singularity.
%

\subsection{Kasner epochs and alternations}
\subsubsection{Anisotropic Kasner universe}
We first consider the geometry near the spacellike singularity as $z\rightarrow\infty$. Numerical data suggest that in the far interior $z\gg z_H$, ($f$, $\chi$, $u$, $A_t$, $\rho_x$) obey the following law:
\begin{equation}\label{asymptotic behavior of fields}
\begin{split}
u\sim z^{n_u},\quad f\sim z^{n_f},\quad\chi\sim n_\chi\ln{z}+\chi_o,\quad \rho_x\sim z^{n_\rho}+\rho_o,\quad A_t\sim z^{n_A}+A_o\,,
\end{split}
\end{equation}
with $n_f$, $n_\chi$, $n_u$, $n_\rho$, $n_A$, $\chi_o$, $\rho_o$ and $A_o$  all constants. Meanwhile, the right hand side of the equations of motion~\eqref{eoms} can be ignored, and then we obtain
\begin{equation}\label{effective equations}
\begin{split}
\left(\frac{e^{\chi/2}\sqrt{u}A_t'}{z^{d-2}}\right)'=0,\quad
\left(\frac{e^{-\chi/2}f\rho_x'}{z^{d-2}\sqrt{u}}\right)'=0,\quad
u''+\left(\frac{f'}{f}-\frac{u'}{2u}-\frac{\chi'}{2}-\frac{d}{z}\right)u'=0,\\
\frac{d}{2}\chi'-d\frac{f'}{f}-\frac{d}{2}\frac{u'}{u}+\frac{d(d+1)}{z}=0,\quad
\left(\frac{d}{z}+\frac{u'}{2u}\right)\frac{f'}{f}-\frac{u'}{2u}\chi'-\frac{d(d+1)}{z^2}=0\,.
\end{split}
\end{equation}
Substituting~\eqref{asymptotic behavior of fields} into~\eqref{effective equations}, we arrive at 
\begin{equation}\label{Relationship}
	\begin{split}
n_\rho=n_u-2,\quad n_A=d-1-\frac{(d-1)n_u}{2d-n_u}\,,\\
 n_f=-n_u+\frac{2d(d+1)-2n_u}{2d-n_u},\quad n_\chi=-n_u+\frac{(2d-2)n_u}{2d-n_u}\,.
	\end{split}
\end{equation}
Changing the $z$ coordinate to the proper time $\tau$ via $\tau\sim z^{-n_f/2}$, we have
\begin{equation}\label{Kasner form}
\begin{split}
ds^2=-d\tau^2+c_t\tau^{2p_t}dt^2+c_x\tau^{2p_x}dx^2+c_{y}\tau^{2p_{y}}\sum_{i=1}^{d-1}dy_i^2\,,
\end{split}
\end{equation}
where $c_t$, $c_x$, $c_{y}$ are all constants and
\begin{equation}\label{Kasner factor}
\begin{split}
p_t=\frac{n_\chi-n_f+2}{n_f},\quad p_x=\frac{2-n_u}{n_f},\quad p_{y}=\frac{2}{n_f}\,.
\end{split}
\end{equation}
One can immediately verify from~\eqref{Relationship} that the above exponents obey
\begin{equation}\label{sphere}
\begin{split}
p_t+p_x+(d-1)p_y=1,\quad p_t^2+p_x^2+(d-1)p_y^2=1\,.
\end{split}
\end{equation}
Therefore, the epoch describes an anisotropic Kasner universe.

\begin{figure}[htbp]
		\centering
		\subfigure[$T_{\blacklozenge}/T_c=0.995$]{\label{Kasner1}
			\includegraphics[width=2.6in]{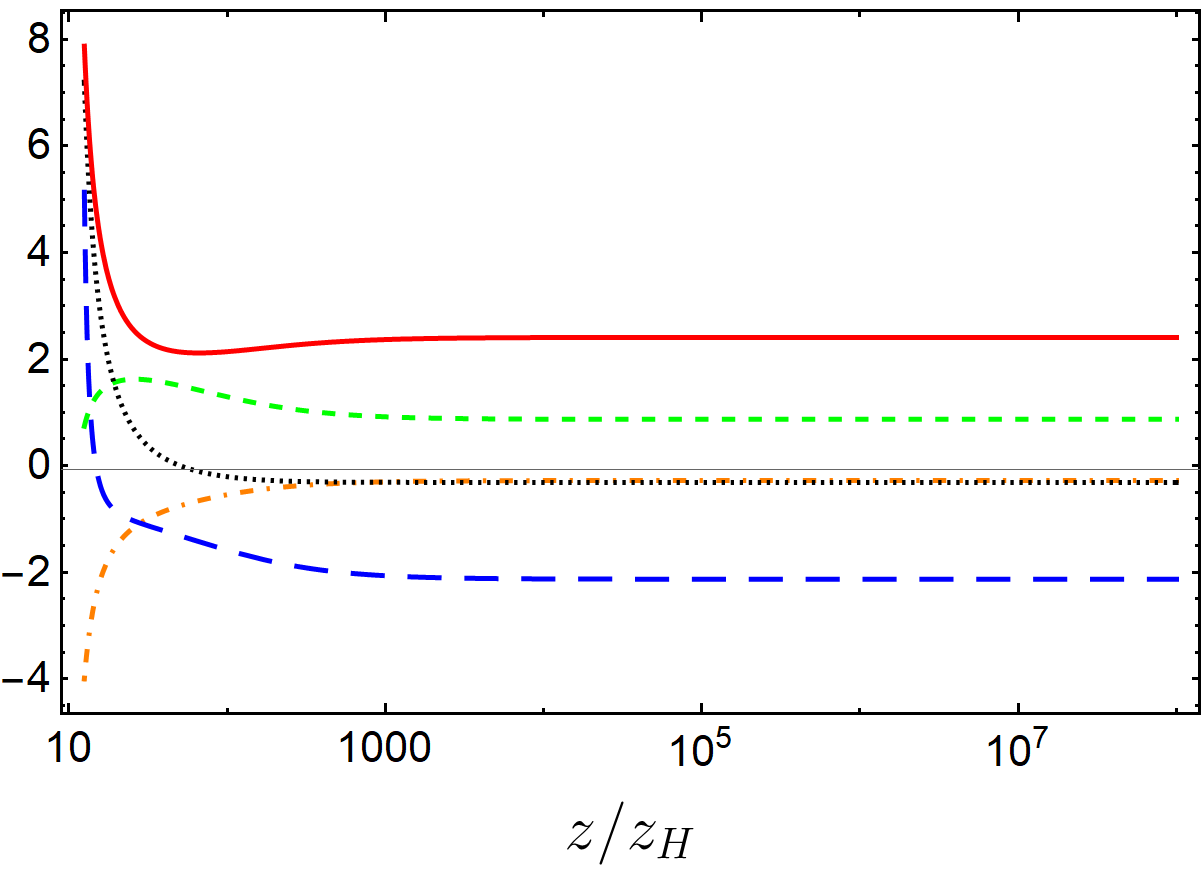}
		}
		\subfigure[$T_{\bullet}/T_c=0.353$]{\label{Kasner3}
			\includegraphics[width=2.6in]{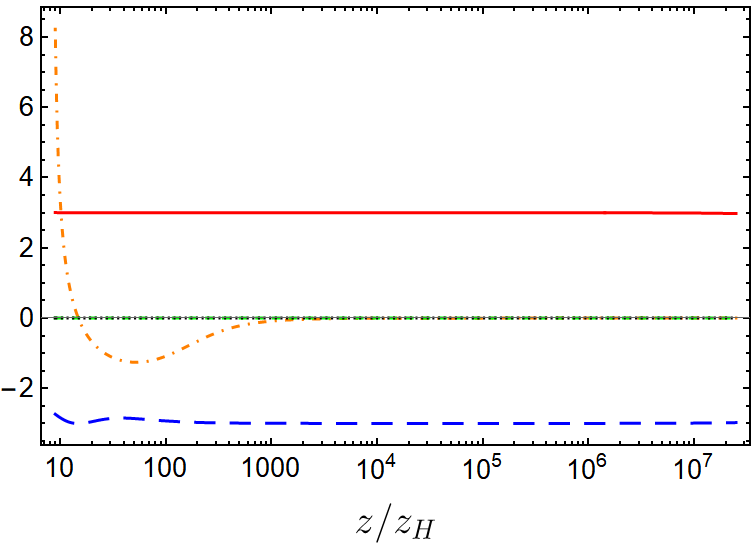}
		}
		\quad
		\subfigure[$T_{\blacktriangle}/T_c=0.684$]{\label{Kasner2}
			\includegraphics[width=4.0in]{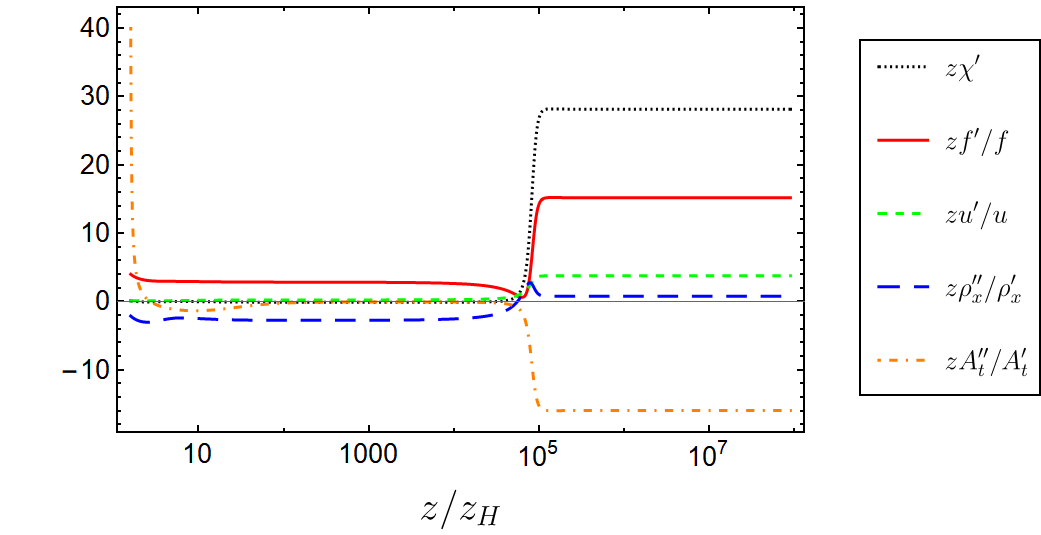}
		}
		\caption{Numerical solutions of various functions inside the black hole event horizon at different temperatures. They all tend to be some constants at large $z/z_H$ and these constants will be determined by the Kasner exponents~\eqref{Relationship}. This is the case for $d=2$, $m^2=0$ and $q=3/2$.}\label{Kasner}
\end{figure}
To confirm our analytic discussion, in Figure.~\ref{Kasner}, we present the behaviors of various functions approaching the singularity for different temperatures, which can be considered as holographic flows that interpolate from a UV radial scaling to a timelike scaling towards late time singularity behind the event horizon. As expected, all curves approach to constant values at large $z/z_H$. Various exponents $n_i (i=u, f, \chi, \rho, A)$ of~\eqref{asymptotic behavior of fields} can be extracted by computing $(zu'/u,\ zf'/f,\ z\chi',\  z\rho_x''/\rho_x',\ zA_t''/A_t')$ from numerical solutions. On the other hand, for a given $n_u$ that is obtained numerically, all other exponents can be obtained analytically from~\eqref{Relationship}. Our comparison between numerical and analytic approaches for various exponents is presented in Table~\ref{NAR}. They match with each other perfectly, verifying the existence of an interior Kasner universe.
\begin{table}[htbp]
	\centering
		\begin{tabular}{c|l|c|c}
				\hline
				\multicolumn{2}{c|}{}&Numerical value&Analytic value\\\hline
				\multirow{4}{*}{\makecell[c]{$T_{\blacklozenge}/T_c=0.995$\\$n_u=1.122774$}}&$n_f$&2.267454&2.267454\\
				&$n_\chi$&-0.342318&-0.342318\\
				&$n_\rho$&-0.877226&-0.877226\\
				&$n_A$&0.609772&0.609772\\
				\hline\hline
				\multirow{4}{*}{\makecell[c]{$T_{\blacktriangle}/T_c=0.684$\\$n_u=3.763766$}}&$n_f$&15.168726&15.168607\\
				&$n_\chi$&28.101218&28.100980\\
				&$n_\rho$&1.763770&1.763766\\
				&$n_A$&-14.932492&-14.932373\\
					\hline\hline
				\multirow{4}{*}{\makecell[c]{$T_{\bullet}/T_c=0.353$\\$n_u=1.26405\times10^{-4}$}}&$n_f$&2.999814&2.999905\\
				&$n_\chi$&$-6.30128\times10^{-5}$&$-6.32006\times10^{-5}$\\
				&$n_\rho$&-1.998342&-1.999874\\
				&$n_A$&0.999473&0.999968\\
				\hline
		\end{tabular}
\tabcaption{Comparison between numerical and analytic approaches of $n_i(i=f, \chi, \rho, A)$ for the cases in Figure.~\ref{Kasner}. They match with each other perfectly.}\label{NAR}
\end{table}
%

\subsubsection{Alternation of  Kasner epochs}
The approximate solution~\eqref{Kasner form} is obtained by neglecting the contribution of matter fields as well as the cosmological constant. To have a stable Kasner epoch, we should require that there are no ``dangerous" terms in the Einstein's equation. We rewrite the last three equations of~\eqref{eoms} as follows:
\begin{equation}\label{metriceoms}
	\begin{split}
		z^2\frac{u''}{u}+\left(\frac{zf'}{f}-\frac{zu'}{2u}-\frac{z\chi'}{2}-d\right)\frac{zu'}{u}=-\frac{2z^4\rho_x'^2}{u}-\frac{2m^2z^2\rho_x^2}{fu}+\frac{2q^2z^4e^{\chi}\rho_x^2A_t^2}{f^2u}\,,\\
		\frac{d}{2}z\chi'-d\frac{zf'}{f}-\frac{d}{2}\frac{zu'}{u}+d(d+1)=\frac{z^4\rho_x'^2}{u}-\frac{2\Lambda}{f}-\frac{z^4e^{\chi} A_t'^2}{2f}-\frac{q^2z^4e^{\chi}\rho_x^2A_t^2}{f^2u}\,,\\
		\left(d+\frac{zu'}{2u}\right)\frac{zf'}{f}-\frac{zu'}{2u}z\chi'-d(d+1)=-\frac{z^4\rho_x'^2}{u}+\frac{2\Lambda}{f}-\frac{m^2z^2\rho_x^2}{fu}+\frac{z^4e^{\chi}A_t'^2}{2f}+\frac{3q^2z^4e^{\chi}\rho_x^2A_t^2}{f^2u}\,.
	\end{split}
\end{equation}
At a Kasner epoch, the left hand side of~\eqref{metriceoms} vanishes. One can estimate the fractional effect of $T_{\mu\nu}$ from the right hand side of~\eqref{metriceoms}, which includes the following terms:
\begin{equation}\label{stability}
\frac{\Lambda}{f},\quad\frac{m^2z^2\rho_x^2}{fu},\quad\frac{z^4\rho_x'^2}{u},\quad\frac{z^4e^\chi A_t'^2}{f},\quad\frac{q^2z^4e^\chi\rho_x^2A_t^2}{f^2u}\,.
\end{equation}
Therefore, one gets as ``dangerous" terms when $z/z_H\rightarrow\infty$ a sum of the above terms. More precisely, we need to check if there exists an open region of the Kasner sphere where all the terms of~\eqref{stability} quickly decays towards the singularity. For a given Kasner universe, all the terms of~\eqref{stability} at large $z/z_H$ behave as a power law, and thus the Kasner epoch will be destroyed if the power is positive. The ``dangerous" terms with  a positive power exponent are summarized in Table~\ref{MTR}.  We find that the  ``dangerous" terms can always appear in some open regions of the Kasner sphere~\eqref{sphere}.  Note that~\eqref{Relationship} implies $n_u\neq2d$.
\begin{table}[h]
	\centering
	\begin{tabular}{c|c|c}
		\hline
		Term&$d$&Range of $n_u$ for positive power\\\hline
		$\nicefrac{\Lambda}{f}$&$d\ge2$&$ (2d,+\infty)$\\\hline
		$\nicefrac{m^2z^2\rho_x^2}{fu}$&$d\ge2$&$ (2d,+\infty)$\\\hline
		$\nicefrac{z^4\rho_x'^2}{u}$&$d\ge2$&$ (2,2d)\cup(2d,+\infty)$\\\hline
		$\nicefrac{z^4e^\chi A_t'^2}{f}$&$d\ge2$&$(-\infty,d)\cup (2d,+\infty)$\\\hline
        \multirow{3}{*}{\makecell[c]{$\nicefrac{q^2z^4e^\chi\rho_x^2A_t^2}{f^2u}$}}&$2\le d\le9$&$(-\infty,0)\cup(d+1,2d)\cup(2d,+\infty)$\\\cline{2-3}
         &\multirow{2}{*}{$d>9$}&$(-\infty,0)\cup(d+1,2d)\cup(2d,+\infty)$\\
         &&$\cup(\frac{d+3-\sqrt{d^2-10d+9}}{2},\frac{d+3+\sqrt{d^2-10d+9}}{2})$\\
    	\hline
\end{tabular}
\caption{We estimate the order of the terms on the right side of~\eqref{metriceoms} on the Kasner universe~\eqref{Kasner form}. Those terms at large $z/z_H$ all behave as a power law.  We give the range of $n_u$ in which the corresponding term yields a positive power exponent and thus violates the Kasner universe.}\label{MTR}
\end{table}

\begin{figure}[htp]
		\centering
		\subfigure[]{\label{kasners}
			\includegraphics[width=2.8in]{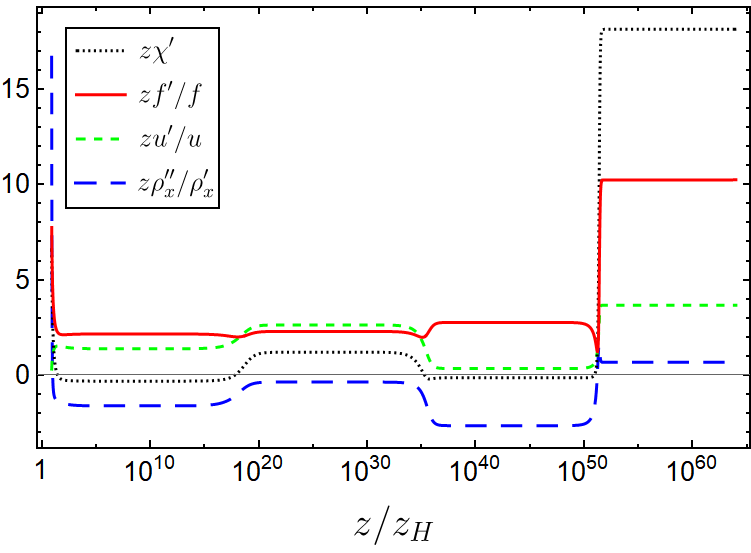}
		}
		\subfigure[]{\label{Rt}
			\includegraphics[width=2.81in]{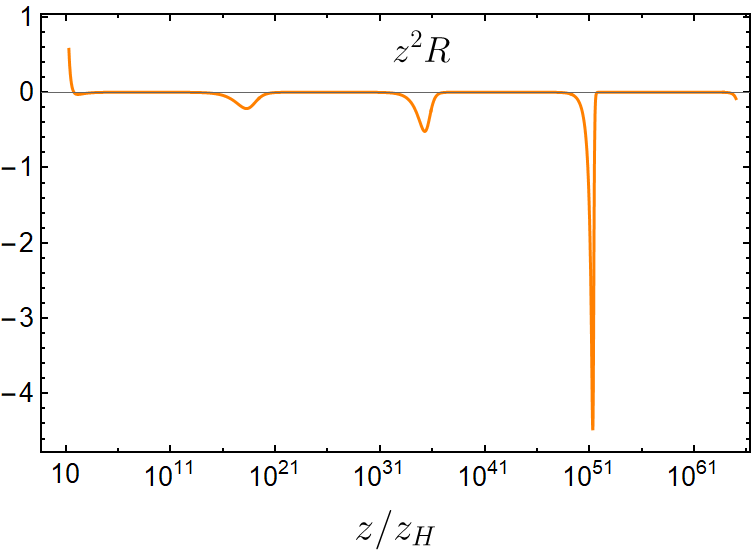}
		}
\caption{Kasner epochs at $T/T_c=0.983$. Four Kasner epochs in the interior are shown in (a). The contribution from the right terms of the last equation of~\eqref{eoms}, denoted as $R$, is plotted in (b). While $R$ is negligible in each Kasner region, it does become important in the transition region between adjacent Kasner epochs. Here $d=2$, $m^2=0$ and $q=3/2$.}\label{Kasners}
\end{figure}

Therefore, the generic solution of the hairy black hole can never reach a monotonic Kasner-like behavior all the way to the singularity at $z\rightarrow\infty$. Actually, we verify numerically (and validate a posteriori) that at the beginning of the Kasner epoch, all the terms of~\eqref{stability} to the equations of motion can be neglected. Nevertheless, the contributions of these terms become important to the background geometry as time ``$z$" evolves and cause a transition to a different Kasner solution with new exponents. A benchmark example is given in Figure.~\ref{Kasners} where four Kasner epochs in the interior are found numerically. To see the effect of matter fields clearly, in the right panel of Figure.~\ref{Kasners}, we plot the right hand of the last equation of~\eqref{eoms} as all terms of~\eqref{stability} are included there. It is clear that in each Kasner epoch  the contribution from matter  fields is negligible, while the matter fields become important in the transition region between adjacent Kasner epochs. The exponents of four Kasner epochs are summarized in Table~\ref{tab:4Kasners}.
%
\begin{table}[htp]
\centering
\begin{tabular}{c|c|c|c|c}
		\hline
		Kasner epochs $(z/z_H)$&$10^5\sim10^{18}$&$10^{18}\sim10^{34}$&$10^{34}\sim10^{51}$&$10^{51}\sim10^{64}$\\
		\hline
		$n_A$&0.474604&{\color{red}-0.908835}&{\color{red}0.908599}&-9.903578\\
		\hline
		$n_u$&{\color{blue}1.375555}&{\color{blue}2.624619}&{\color{blue}0.336036}&{\color{blue}3.663967}\\
		\hline
		Sum&\multicolumn{2}{c|}{{\color{blue}4.000168}}&\multicolumn{2}{c}{{\color{blue}4.000002}}\\
		\hline
	\end{tabular}
	\tabcaption{Exponents of Kasner epochs at $T/T_c=0.983$ by fitting the solution of Figure.~\ref{Kasners}. The sum of $n_u$ of the first two Kasner epochs and the sum of the third and the fourth Kasner epochs are equal to 4 (blue). The values of $n_A$ at the second and third Kasner epochs are approximately equal, but with opposite signs (red).}\label{tab:4Kasners}
\end{table}

In order to see more times of alternation of the Kasner epochs, one needs to solve the equations of motion~\eqref{eoms} to sufficiently large $z/z_H$. Unfortunately, due to the limitation of computing power, we are not able to do this.\,\footnote{In practice, we find that the equations of motion become increasingly stiff as $z/z_H$ increases. As evident from Figure.~\ref{Kasners}, the system includes some terms that lead to rapid variation in the solution, making the equations numerically unstable.} Nevertheless, our analysis suggests that near the spacelike singularity there exhibits never ending oscillatory behavior\,\footnote{Th chaotic behavior inside the static, spherically symmetric black holes of the Einstein-Yang-Mills theory was studied in~\cite{Donets:1996ja,Breitenlohner:1997hm}}, and we do find some interesting rule defining the change of Kasner exponents from one epoch to the next. In the present work, we focus on $d=2$, for which our numerics suggest the following features (see Table~\ref{tab:4Kasners} and Table~\ref{tab:kasner} for various numerical examples).
\begin{itemize}
  \item (a) For $n_u<2$, the dominant term that triggers the alternation of Kasner epochs is $z^4e^\chi A_t'^2/f$. The sum of $n_u$ and $\tilde{n}_u$ at adjacent Kasner epochs is found to be
\begin{equation}\label{rule1}
n_u+\tilde{n}_u=4\,.
\end{equation}

  \item (b) For $2<n_u<4$, the dominant term that triggers the alternation of Kasner epochs is $z^4\rho_x'^2/u$. It turns out that $n_A$ and $\tilde{n}_A$ at adjacent Kasner epochs are opposite values, \emph{i.e.}
\begin{equation}\label{rule2}
n_A=-\tilde{n}_A\,.
\end{equation}
  \item (c) For $n_u>4$, all five terms of~\eqref{stability} will become important to the background for sufficiently large $z/z_H$, for which the dominant term is $\frac{q^2z^4e^\chi\rho_x^2A_t^2}{f^2u}$. However, for all parameters we have considered, we do not find any Kasner universe that has $n_u>4$.
\end{itemize}
\begin{table}
	\centering
	\begin{tabular}{c|c|c|c|c}
		\hline
		&Kasner epochs $(z/z_H)$&$n_u$&$n_A$&Sum\\\hline
		\multirow{3}{*}{\makecell[c]{$T/T_c=0.998$}}&$10^3\sim10^{146}$&{\color{blue}1.173335}&0.584905&\multirow{2}{*}{\makecell[c]{{\color{blue}4.000000}}}\\\cline{2-4}
		&$10^{146}\sim10^{290}$&{\color{blue}2.826665}&{\color{red}-1.409087}&\\\cline{2-5}
		&$10^{290}\sim10^{436}$&-2.769187&{\color{red}1.409087}&\\
		\hline\hline
		\multirow{3}{*}{\makecell[c]{$T/T_c=0.997$}}&$10^2\sim10^{127}$&{\color{blue}1.240325}&0.550554&\multirow{2}{*}{\makecell[c]{{\color{blue}4.000000}}}\\\cline{2-4}
		&$10^{127}\sim10^{250}$&{\color{blue}2.759675}&{\color{red}-1.224960}&\\\cline{2-5}
		&$10^{250}\sim10^{380}$&-1.160680&{\color{red}1.224893}&\\
		\hline\hline
		\multirow{3}{*}{\makecell[c]{$T/T_c=0.994$}}&$10^3\sim10^{48}$&{\color{blue}1.218098}&0.562135&\multirow{2}{*}{\makecell[c]{{\color{blue}4.000000}}}\\\cline{2-4}
		&$10^{48}\sim10^{95}$&{\color{blue}2.781902}&{\color{red}-1.283808}&\\\cline{2-5}
		&$10^{95}\sim10^{140}$&-1.585010&{\color{red}1.283793}&\\
		\hline\hline
		\multirow{3}{*}{\makecell[c]{$T/T_c=0.991$}}&$10^3\sim10^{32}$&{\color{blue}1.122742}&0.609818&\multirow{2}{*}{\makecell[c]{{\color{blue}3.999968}}}\\\cline{2-4}
		&$10^{32}\sim10^{64}$&{\color{blue}2.877226}&{\color{red}-1.562605}&\\\cline{2-5}
		&$10^{64}\sim10^{90}$&-5.145021&{\color{red}1.562602}&\\
		\hline\hline
		\multirow{3}{*}{\makecell[c]{$T/T_c=0.985$}}&$10^2\sim10^{16}$&{\color{blue}1.014178}&0.660504&\multirow{2}{*}{\makecell[c]{{\color{blue}4.000206}}}\\\cline{2-4}
		&$10^{16}\sim10^{32}$&{\color{blue}2.986028}&{\color{red}-1.944810}&\\\cline{2-5}
		&$10^{32}\sim10^{45}$&-68.574549&{\color{red}1.944884}&\\
		\hline\hline
	\end{tabular}
	\tabcaption{The values $n_u$ and $n_A$ at three Kasner epochs for five different temperatures. The sum of $n_u$ of the first and second Kasner epochs are, within numerical error, equal to 4 (blue), while the values of $n_A$ at the second and third Kasner epochs are opposite values (red). Here $d=2$, $m^2=0$ and $q=3/2$.}\label{tab:kasner}
\end{table}

In contrast to the BKL analysis where the effect of the spatial derivatives of metric plays the key role, our Kasner alternations are induced by the vector condensate and the U(1) gauge field. Nevertheless, we find that the power substitution rule of the adjacent Kasner epochs follows the BKL type~\cite{Belinsky:1970ew}. Specifically, for case $(a)$ triggered by the kinetic term of gauge potential $z^4e^\chi A_t'^2/f$, the negative power of $\tau$ in~\eqref{Kasner form} flips from the $t$ direction to the $x$ direction:

\begin{equation}\label{BKL1}
	p_t\rightarrow-\frac{p_t}{1+2p_t},\quad p_x\rightarrow\frac{2p_t+p_x}{1+2p_t},\quad p_y\rightarrow\frac{2p_t+p_y}{1+2p_t}\,.
\end{equation}
In terms of the exponent $n_u$, the above rule corresponds to $n_u\rightarrow4-n_u$, \emph{i.e.} the rule in~\eqref{rule1}. For case $(b)$ driven by the kinetic term of vector $z^4\rho_x'^2/u$, the negative power of $\tau$ flips from the $x$ direction to the $t$ direction with the following change on $p_i(i=t, x, y)$:
\begin{equation}\label{BKL2}
	p_t\rightarrow\frac{2p_x+p_t}{1+2p_x},\quad p_x\rightarrow-\frac{p_x}{1+2p_x},\quad p_y\rightarrow\frac{2p_x+p_y}{1+2p_x}\,.
\end{equation}
In terms of the exponent $n_A$, such rule means $n_A\rightarrow-n_A$, \emph{i.e.} the rule in~\eqref{rule2}. In both cases, the Kasner exponent $p_y$ of~\eqref{Kasner form} is always positive.

\begin{figure}
	\centering
	\subfigure[$T_{\lozenge}/T_c=0.991$]{\label{Tc1}
		\includegraphics[width=2.8in]{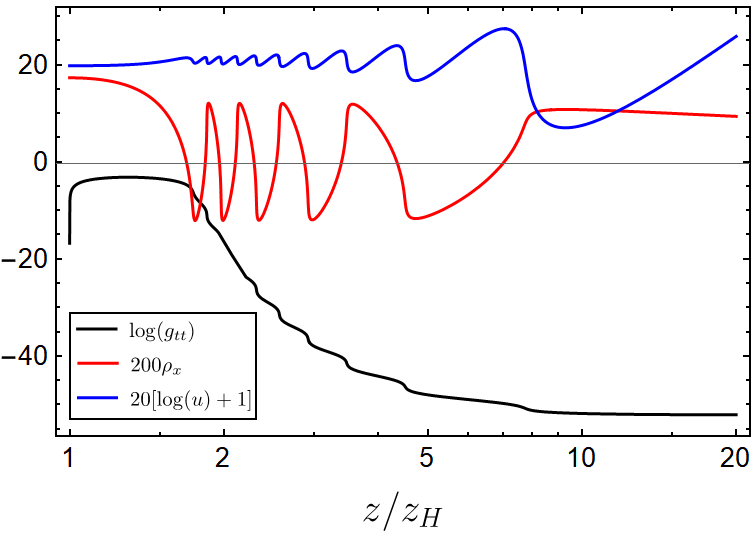}
	}
	\subfigure[$T_{\vartriangle}/T_c=0.966$]{\label{Tc2}
		\includegraphics[width=2.8in]{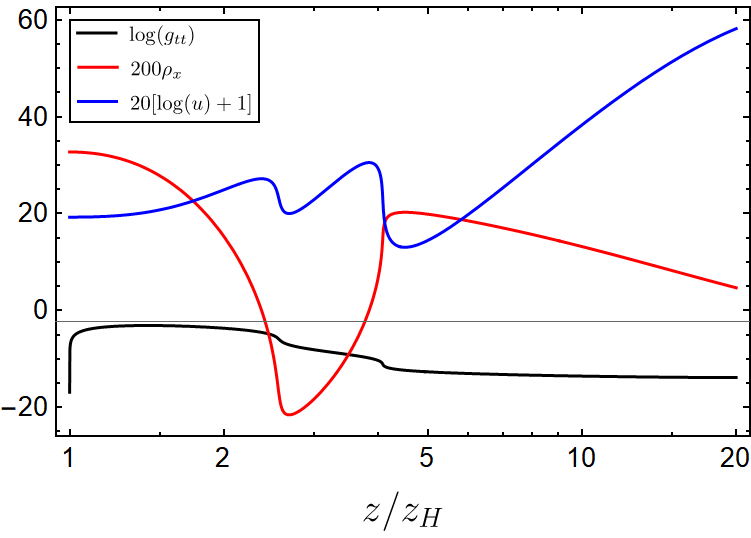}
	}
	\caption{Nonlinear dynamics of the hairy black hole near the would-be inner horizon of AdS RN at $T/T_c=0.991$ (a) and $T/T_c=0.966$ (b). $g_{tt}$ experiences rapid collapse towards the singularity and $\rho_x$ oscillates. Moreover, the anisotropic factor $u(z)$ also oscillates and has a period that is half of that of the vector condensate. The imprint of the oscillations on the metric is manifest. These nonlinear effects become less dramatic at lower temperatures. The plot is  the case for $d=2$, $m=0$ and $q=3/2$.}\label{Tc}
\end{figure}
%
%
\subsection{Collapse of ER bridge and Josephson oscillations}
We now consider  the nonlinear dynamics before Kasner epochs. As the temperature is above $T_c$, the black hole solution is AdS-RN~\eqref{RN} which has an inner Cauchy horizon. However, the vector hair will develop spontaneously below $T_c$. Our poof in Section~\ref{sec:proof} suggests that  an arbitrarily small amount of vector hair will result in an instability of  the inner Cauchy horizon of the black hole and will make the inner horizon disappear.

In Figure.~\ref{Tc}, we show the nonlinear dynamics that occur near the would-be inner horizon due to the instability of the Cauchy horizon triggered by vector hairs. Slightly below $T_c$, it is manifest from Figure.~\ref{Tc1} that near the would-be inner horizon, $g_{tt}$ experiences a rapid collapse, meanwhile the charged vector $\rho_x$ oscillates. The former is known as collapse of the ER bridge, while for the latter we call it Josephson oscillation following the spirit of charged scalar case~\cite{Hartnoll:2020fhc}. Different from the scalar case~\cite{Hartnoll:2020rwq,Hartnoll:2020fhc,An:2021plu}, oscillation also happens in $g_{xx}$. More precisely, $u(z)=z^2g_{xx}$, which characterizes the anisotropy of the geometry, oscillates with twice the frequency of the vector hair $\rho_x$.  Note however that, as shown in Figure.~\ref{Tc2}, such non-linear dynamics will become less dramatic as the temperature is decreased.

\begin{figure}
	\centering
	\subfigure[]{\label{fit1}
		\includegraphics[width=2.86in]{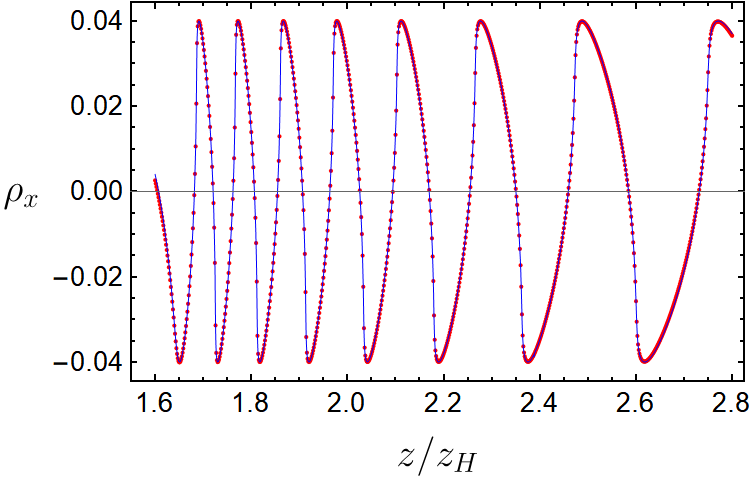}
	}
	\subfigure[]{\label{fit2}
		\includegraphics[width=2.76in]{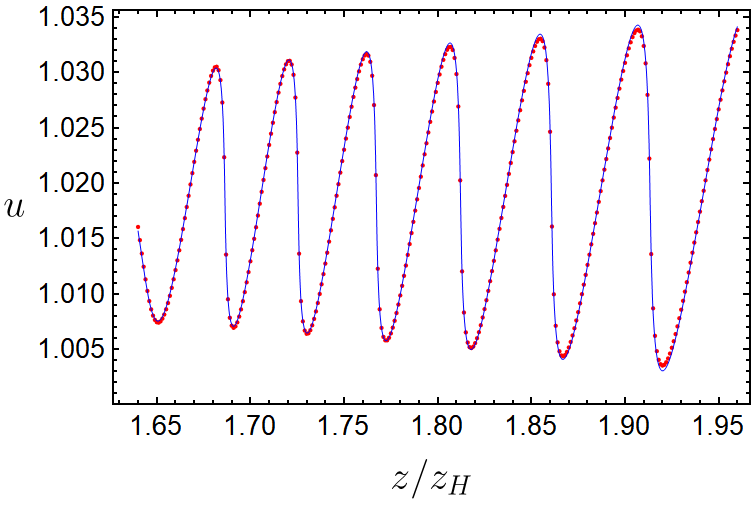}
	}
	\caption{Comparison of the numerical solutions (red dotted lines) and fits to the analytic expressions~\eqref{rho} and~\eqref{u} (solid blue curves) describing Josephson oscillations. The resulting fitting parameters are  $\rho_o=-0.039996, \Omega=0.59135, \varphi_o=-4.3626,\ u_o=1.0290$. This is the case with $d=2$, $m=0$, $q=3/2$, $T/T_c=0.996$ and $z_{\star}/z_H=9/5$.}\label{fit}
\end{figure}
Because of the complication of equations of motion~\eqref{eoms}, it is difficult to obtain analytic solutions for the full nonlinear regimes of Figure.~\ref{Tc}. Nevertheless, we are able to find good analytic approximations near the would-be inner horizon of AdS-RN closed $T_c$. Let's consider the equation of motion of $\rho_x$:
\begin{equation}\label{eomrho}
	\begin{split}
		\frac{e^{-\chi/2}f}{\sqrt{u}}\left(\frac{e^{-\chi/2}f\rho_x'}{\sqrt{u}}\right)'=-\frac{q^2A_t^2}{u}\rho_x\,,
	\end{split}
\end{equation}
where we have chosen $d=2$ and $m^2=0$. As guided by numerical exploration, in the oscillation regime close to the would-be inner horizon, we can set $A_t^2/u=\Omega^2$ a constant. Then, we obtain from~\eqref{eomrho} that
\begin{equation}\label{rho}
	\begin{split}
		\rho_x=\rho_o\cos\left(q\Omega\int_{z_\star}^{z}\frac{e^{\chi/2}\sqrt{u}dy}{f}+\varphi_o\right)\approx\rho_o\cos\left(q\Omega\int_{z_\star}^{z}\frac{e^{\chi/2}dy}{f}+\varphi_o\right)\,,
	\end{split}
\end{equation}
where $z_\star$, $\rho_o$ and $\varphi_o$ are constants, and we have used the fact that $u$ can be approximated to be 1 in the integrand for small values of the vector field at the horizon, \emph{i.e.} as $T\rightarrow T_c$. As verified in Figure.~\ref{fit1}, numerical solutions to the equations of motion are found to be well fitted by~\eqref{rho} at Josephson oscillations.

Moreover, using the conserved charge~\eqref{charge2} and considering that $f$ vanishes at the event horizon, we can obtain that
\begin{equation}
u'+2z^2\rho_x\rho_x'=0\,.
\end{equation}
Then, $u$ can be solved explicitly as
\begin{equation}\label{u}
u(z)=-\int_{z_\star}^{z}y^2\frac{d\rho_x^2}{dy} dy+u_o\,,
\end{equation}
with $u_o$ a constant. In the oscillation regime, \eqref{rho} should be inserted into the integral of~\eqref{u}. It is now clear that $u$ is driven by $(\rho_x^2)'$, thus oscillating with twice the frequency of $\rho_x$. The form~\eqref{u} is verified by fitting the numerical solutions to the full equations of motion~\eqref{eoms}, see Figure.~\ref{fit2}.

\section{Conclusion and Discussion}\label{sec:discussion}
We have considered the interior structure of anisotropic black holes with charged vector hairs. After showing how to construct the Noether charge with higher derivatives, we have proven that there is no inner Cauchy horizon for those hairy black holes.  Although we have limited ourselves to the free vector field case with charge $q$ and mass $m^2$~\eqref{action}, our proof can be straightforwardly extended to quite generic cases. A natural extension is given as follows.
\begin{equation}\label{actionnew}
\mathcal{L}_m^{\text{ext}}=-\frac{X(\rho_\mu^\dagger\rho^\mu, |\rho_\mu\rho^\mu|^2)}{4}F_{\mu\nu} F^{\mu \nu}-\frac{1}{2}\rho_{\mu\nu}^\dagger\rho^{\mu\nu}-V(\rho_\mu^\dagger\rho^\mu, |\rho_\mu\rho^\mu|^2)+iq\gamma \rho_\mu\rho_\nu^\dagger F^{\mu\nu},
\end{equation}
where $X$ and $V$ are arbitrary smooth functions of $\rho_\mu^\dagger\rho^\mu$ and $|\rho_\mu\rho^\mu|^2$. The conserved charge associated with the generalized theory~\eqref{actionnew} is given by
\begin{equation}
\mathcal{Q}=\frac{e^{\chi/2}\sqrt{u}}{z^{d}}\left[\left(fe^{-\chi}\right)'-z^2XA_tA_t'\right]\,,
\end{equation}
which does not depend on the potential $V$ of~\eqref{actionnew}. Following a similar discussion in Section~\ref{sec:proof}, it is easy to show that the Cauchy horizon for black holes with charged vector hair is not allowed.

One can also consider modified gravity with high derivative corrections. One of the most studied cases is the Gauss-Bonnet term. The action reads
\begin{equation}\label{GB}
S_{GB}=\frac{1}{2\kappa_N^2}\int d^{d+2} x\sqrt{-g}\left[\mathcal{R}+\alpha_{GB}(\mathcal{R}^2-4\mathcal{R}_{\mu\nu}\mathcal{R}^{\mu\nu}+\mathcal{R}_{\mu\nu\gamma\delta}\mathcal{R}^{\mu\nu\gamma\delta})-2\Lambda+\mathcal{L}_m\right]\,.
\end{equation}
One can check that the resulting effective action of the hairy black hole~\eqref{ansatz} still has the scaling symmetry of~\eqref{scaling1}. Then, we can obtain the Noether charge
\begin{equation}
\mathcal{Q}=\frac{e^{\chi/2}\sqrt{u}}{z^{d}}\left[\left(fe^{-\chi}\right)'-z^2A_tA_t'\right]+2(d-2)\alpha_{GB}\frac{e^{\chi/2}f}{z^{d} \sqrt{u}}(fe^{-\chi})'\left[(1-d)u+zu'\right]\,.
\end{equation}
It is manifest that the contribution from the Gauss-Bonnet term is proportional to the blacking function $f(z)$ which vanishes at a horizon. Therefore, our proof in Section~\ref{sec:proof} applies to the case with the Gauss-Bonnet term. Once again, there is no Cauchy horizon in Einstein-Gauss-Bonnet black holes with vector hair. The construction of the radially conserved charge $\mathcal{Q}$ is easy to generalize to other cases, including the scalar case considered in~\cite{Cai:2020wrp,An:2021plu,Hartnoll:2020fhc}. It is interesting to see if this method can be generalized to inhomogeneous horizon geometries, at least, for the planar horizon topology case.

The hairy black holes approach to a spacelike singularity at late interior time. Just below the critical temperature $T_c$  when the vector field is uniformly small everywhere outside the horizon, the interior dynamics evolve several distinct epochs, including a rapid collapse of the ER bridge, and the Josephson oscillations of the vector condensate which in turn induces the oscillation of the anisotropy of spatial geometry with twice the frequency of the condensate. Then the geometry enters into Kasner epochs with spatial anisotropy. Due to the effects from vector condensate and U(1) gauge potential, we showed that it can never reach a monotonic
Kasner-like behavior. There is generically a never-ending alternation of Kasner epochs towards the singularity. Due to the limitation of computing power, we are  able to see as many as four Kasner epochs as shown in Figure.~\ref{Kasners}. An efficient numerical algorithm is required to solve the equations of motion~\eqref{eoms} for sufficiently large $z/z_H$. Nevertheless, we found that the character of the Kasner epoch alternation with flipping of powers is BKL type~\cite{Belinsky:1970ew}, see~\eqref{BKL1} and~\eqref{BKL2}. It will be interesting to have a deep understanding on the general rule for the change of Kasner exponents from one epoch to the next in the present theory. While we have limited to the discussion of interior dynamics for the vector hair that develops spontaneously, the main features should apply to the hairy black hole with explicit sources.
It would be interesting to extend the discussion to more general cases, such as other spacetime dimensions and new couplings. We hope to report on the results in these directions in the near future.

\section*{Acknowledgements}
This work was partially supported by the National Natural Science Foundation of China Grants No.12122513, No.12075298, No.11991052, No.12047503, No.12005155 and No.11821505, and by the Key Research Program of the Chinese Academy of Sciences (CAS) Grant NO. XDPB15, the CAS Project for Young Scientists in Basic Research YSBR-006 and the Key Research Program of Frontier Sciences of CAS.

\appendix

\section{Einstein-SU(2) Yang-Mills Model}\label{app:YMs}
Another gravity theory with charged vector hair is the Einstein-SU(2) Yang-Mills model~\cite{Gubser:2008wv}:
\begin{equation}
\mathcal{L}_m^{\text{YM}}=-\frac{1}{4\hat{g}^2}F^a_{\mu\nu} F^{a\mu \nu}\,,
\end{equation}
where $\hat{g}$ is the Yang-Mills coupling constant. The gauge field is given by $A=A^a_{\mu}\tau^adx^{\mu}$ with $a,b,c=(1,2,3)$ the indices of the SU(2) group generators
$\tau^a=\sigma^a/2i$ ($\sigma^a$ are Pauli matrices). The Yang-Mills field strength reads
\begin{equation}
 F^a_{\mu\nu}=\partial_\mu A^a_\nu-\partial_\nu A^a_\mu + \epsilon^{abc}A^b_\mu
 A^c_\nu,
\end{equation}
where $\epsilon^{abc}$ is the totally antisymmetric tensor with $\epsilon^{123}=+1$.
There already exists  evidence that the complex vector field model  considered in this work is a generalization of the SU(2) model to the case with a general mass $m^2$ and magnetic moment characterized by $\gamma$~\cite{Cai:2013pda,Cai:2013kaa,Nie:2014qma}. However, the discussions were done only under particular ansatz. Here we present a complete proof.

We identify $A_\mu^3$ as the electromagnetic U(1) gauge field $\hat{A}_\mu$ and introduce a complex vector $\hat{\rho}_\mu$
\begin{equation}
\hat{\rho}_\mu=A_\mu^1-i A_\mu^2\,.
\end{equation}
After some simple algebra, we obtain
\begin{equation}
\begin{split}
F_{\mu\nu}^3&=\hat{F}_{\mu\nu}-i \hat{\rho}_{[\mu}\hat{\rho}^\dagger_{\nu]}\,,\\
F_{\mu\nu}^1-i F_{\mu\nu}^2&=\hat{D}_\mu \hat{\rho}_\nu-\hat{D}_\nu \hat{\rho}_\mu\equiv \hat{\rho}_{\mu\nu}\,,\\
F_{\mu\nu}^1+i F_{\mu\nu}^2&=(\hat{D}_\mu \hat{\rho}_\nu-\hat{D}_\nu \hat{\rho}_\mu)^\dagger= {\hat{\rho}}_{\mu\nu}^\dagger\,,
\end{split}
\end{equation}
with $\hat{F}_{\mu\nu}=\nabla_\mu \hat{A}_\nu-\nabla_\nu \hat{A}_\mu$ and $\hat{D}_\mu=\nabla_\mu-i \hat{A}_\mu$. Then, we have
\begin{equation}
\begin{split}
F^a_{\mu\nu} F^{a\mu \nu}&={\hat{\rho}}_{\mu\nu}^\dagger{\hat{\rho}}^{\mu\nu}+(\hat{F}_{\mu\nu}-i \hat{\rho}_{[\mu}\hat{\rho}^\dagger_{\nu]})(\hat{F}^{\mu\nu}-i {\hat{\rho}}^{[\mu}{\hat{\rho}}^{\dagger\nu]})\,,\\
&={\hat{\rho}}_{\mu\nu}^\dagger{\hat{\rho}}^{\mu\nu}+\hat{F}_{\mu\nu}\hat{F}^{\mu\nu}-2i  \hat{\rho}_{\mu}\hat{\rho}^\dagger_{\nu}\hat{F}^{\mu\nu}+\frac{1}{2}(|\rho_\mu^\dagger\hat{\rho}^\mu|^2-|\hat{\rho}_\mu\hat{\rho}^\mu|^2)\,.
\end{split}
\end{equation}
After making the following scaling transformation
\begin{equation}
\hat{A}_\mu=\hat{g}A_\mu,\quad \hat{\rho}_\mu=\sqrt{2}\hat{g}\rho_\mu\,,
\end{equation}
we arrive at
\begin{equation}\label{actionYM}
\mathcal{L}_m^{\text{YM}}=-\frac{1}{4}F_{\mu\nu} F^{\mu \nu}-\frac{1}{2}\rho_{\mu\nu}^\dagger\rho^{\mu\nu}-\frac{\hat{g}^2}{2}(|\rho_\mu^\dagger\rho^\mu|^2- |\rho_\mu\rho^\mu|^2)+iq\gamma \rho_\mu\rho_\nu^\dagger F^{\mu\nu},
\end{equation}
with $q=\hat{g}$ and $\gamma=1$. Therefore, the SU(2) model is a special case of our complex vector model~\eqref{actionnew} with vanishing mass and gyromagnetic ratio $\gamma=1$. Under ansatz~\eqref{ansatz}, the last two terms of~\eqref{actionYM} can be ignored and the SU(2) model will give the same equations of motion as~\eqref{eoms} with $q=\hat{g}$ and $m^2=0$.

\section{Proof of the Noether charge}\label{app:charge}
We provide proof of the radially conserved Noether charge used in~\eqref{initial charge}.
We begin with the following action where the Lagrangian contains some functions $\mathcal{F}_a$ with variable $z$ in an interval $(z_1, z_2)$.
\begin{equation}\label{action1}
	\begin{split}
		S=\int_{z_1}^{z_2} dz \mathcal{L}(\mathcal{F}_a, \mathcal{F}_a', \mathcal{F}_a''; z)\,,
	\end{split}
\end{equation}
where the index $a$ labels different functions and prime denotes the derivative with respect to $z$. Note that $\mathcal{L}$ depends on the derivatives of $\mathcal{F}_a$ up to the second order.

We consider a continuous symmetry with a real parameter $\epsilon$. For an infinitesimal transformation for which $\epsilon$ is small, we have, to first order in $\epsilon$, that
\begin{equation}\label{symmetry1}
	\begin{split}
		z\rightarrow \tilde{z}=z+ \epsilon Z(z),\quad \mathcal{F}_a(z) \rightarrow \widetilde{\mathcal{F}}_a(\tilde{z})=\mathcal{F}_a(z)+\epsilon F_a(z)\,,
	\end{split}
\end{equation}
from which the transformation of $\mathcal{F}_a'(z)$ and $\mathcal{F}_a''(z)$ can be obtained.
\begin{equation}\label{transformations}
	\begin{split}
	\mathcal{F}_a'(z) \rightarrow \widetilde{\mathcal{F}}_a'(\tilde{z})&=\frac{d\widetilde{\mathcal{F}}_a}{dz}\frac{dz}{d\tilde{z}}=[\mathcal{F}_a'(z)+\epsilon F_a'(z)][1-\epsilon Z'(z)]\\
	&=\mathcal{F}_a'(z)+\epsilon [F_a'(z)-\mathcal{F}_a'(z) Z'(z)]\,,\\
	\mathcal{F}_a''(z) \rightarrow \widetilde{\mathcal{F}}_a''(\tilde{z})&=\frac{d\widetilde{\mathcal{F}}'_a}{dz}\frac{dz}{d\tilde{z}}=\left\{\mathcal{F}_a''(z)+\epsilon \frac{d[F_a'(z)-\mathcal{F}_a'(z) Z'(z)]}{dz}\right\}[1-\epsilon Z'(z)]\\
	&=\mathcal{F}_a''(z)+\epsilon [F_a''(z)-2\mathcal{F}_a''(z) Z'(z)-\mathcal{F}_a'(z) Z''(z)]\,.\\
	\end{split}
\end{equation}
Then we can compute the variation of $S$ to first order in $\epsilon$.
\begin{equation}\label{variation}
	\begin{split}
	\delta_\epsilon S&=\int_{z_1}^{z_2} \delta(dz) \mathcal{L}(\mathcal{F}_a, \mathcal{F}_a', \mathcal{F}_a''; z) + dz\delta \mathcal{L}(\mathcal{F}_a, \mathcal{F}_a', \mathcal{F}_a''; z)\\
	&=\int_{z_1}^{z_2} d(\delta z)\mathcal{L} +\frac{\partial\mathcal{L}}{\partial z}\delta z + \frac{\partial\mathcal{L}}{\partial\mathcal{F}_a}\delta \mathcal{F}_a +\frac{\partial\mathcal{L}}{\partial\mathcal{F}'_a}\delta \mathcal{F}'_a +\frac{\partial\mathcal{L}}{\partial\mathcal{F}''_a}\delta \mathcal{F}''_a\,, \\
	&=\epsilon \int_{z_1}^{z_2} dz Z'\mathcal{L} +\frac{\partial\mathcal{L}}{\partial z}Z + \frac{\partial\mathcal{L}}{\partial\mathcal{F}_a} F_a +\frac{\partial\mathcal{L}}{\partial\mathcal{F}'_a}(F_a'-\mathcal{F}_a' Z')+\frac{\partial\mathcal{L}}{\partial\mathcal{F}''_a}(F_a''-2\mathcal{F}_a'' Z'-\mathcal{F}_a' Z'')\,, \\
	&=\epsilon \int_{z_1}^{z_2} dz Z'\mathcal{L} +\frac{d\mathcal{L}}{d z}Z + \frac{\partial\mathcal{L}}{\partial\mathcal{F}_a}(F_a-Z\mathcal{F}_a') +\frac{\partial\mathcal{L}}{\partial\mathcal{F}'_a}\frac{d(F_a-Z\mathcal{F}_a')}{dz}+\frac{\partial\mathcal{L}}{\partial\mathcal{F}''_a}\frac{d^2(F_a-Z\mathcal{F}_a')}{dz^2}\,,\\
	&=\epsilon \left\{ Z\mathcal{L}+ \left[\frac{\partial\mathcal{L}}{\partial\mathcal{F}_a'}-\frac{d}{dz}\left(\frac{\partial\mathcal{L}}{\partial\mathcal{F}_a''}\right)\right](F_a-Z\mathcal{F}_a')+\left(\frac{\partial\mathcal{L}}{\partial\mathcal{F}_a''}\right)\frac{d(F_a-Z\mathcal{F}_a')}{dz}\right\} \bigg|_{z_1}^{z_2}\\
	&\quad + \epsilon \int_{z_1}^{z_2} dz \left[\frac{\partial\mathcal{L}}{\partial\mathcal{F}_a}-\frac{d}{dz}\left(\frac{\partial\mathcal{L}}{\partial\mathcal{F}_a'}\right)+\frac{d^2}{dz^2}\left(\frac{\partial\mathcal{L}}{\partial\mathcal{F}_a''}\right)\right](F_a-Z\mathcal{F}_a')\,,
	\end{split}
\end{equation}
where the Einstein summation convention for the index $a$ has been used.

Considering the Euler-Lagrange equations,
\begin{equation}\label{EL}
	\frac{\delta S}{\delta \mathcal{F}_a}=\frac{\partial\mathcal{L}}{\partial\mathcal{F}_a}-\frac{d}{dz}\left(\frac{\partial\mathcal{L}}{\partial\mathcal{F}_a'}\right)+\frac{d^2}{dz^2}\left(\frac{\partial\mathcal{L}}{\partial\mathcal{F}_a''}\right)=0\,,
\end{equation}
and $\delta_\epsilon S=0$ as a symmetry of the action~\eqref{action1}, by definition,  we immediately obtain that
\begin{equation}\label{Noether charge}
	\begin{split}
		\mathcal{Q}_{\text{Noether}}= Z\mathcal{L}+ \left[\frac{\partial\mathcal{L}}{\partial\mathcal{F}_a'}-\frac{d}{dz}\left(\frac{\partial\mathcal{L}}{\partial\mathcal{F}_a''}\right)\right](F_a-Z\mathcal{F}_a')+\left(\frac{\partial\mathcal{L}}{\partial\mathcal{F}_a''}\right)\frac{d(F_a-Z\mathcal{F}_a')}{dz}\,,
	\end{split}
\end{equation}
is independent of $z$ for solutions to the Euler-Lagrange equations~\eqref{EL}, \emph{i.e.} $\mathcal{Q}_{\text{Noether}}'=0$. Therefore, $\mathcal{Q}_{\text{Noether}}$ is the analog of Noether charge associated with the symmetry~\eqref{symmetry1}. Here we have considered the case up to the second order derivatives, but the generalization to the higher order derivatives is straightforward.

\section{Proof for $A_t=0$ at a horizon}\label{app:At}

In this section we will show that for the anisotropic black holes with charged vector hairs (\emph{i.e.} $\rho_x$ is not zero somewhere), $A_t$ must be zero at a horizon $z=z_i$ where $f(z_i)$ vanishes.

For the hairy black hole with ansatz~\eqref{ansatz}, the equation of motion~\eqref{vector} for the charged vector field reads
\begin{equation}\label{vectoreom}
	\begin{split}
z^{d-2}e^{\chi/2}\sqrt{u}\left(\frac{e^{-\chi/2}f\rho_x'}{z^{d-2}\sqrt{u}}\right)'=\frac{m^2\rho_x}{z^{2}}-\frac{q^2e^{\chi}\rho_xA_t^2}{f}\,.
   \end{split}
\end{equation}
Note that the denominator of the last term in the above equation vanishes at $z_i$. The smoothness of the spacetime geometry near $z_i$ then yields
\begin{center}
	(a) $A_t(z_i)=0$, \qquad or\qquad (b) $A_t(z_i)\neq0$,\;\;$\rho_x(z_i)=0$\,.
\end{center}
What we shall do next is to exclude the second case. Let's first assume that case (b) is true. Since $\rho_x(z)$ is smooth and nonzero somewhere, one has the following Taylor expansion:
\begin{equation}\label{tayrho}
	\rho_x=\rho_n(z-z_i)^n+\rho_{n+1}(z-z_i)^{n+1}+\cdots,~~n\geq1\,,
\end{equation}
with $\rho_n\neq0$. Similarly, for $f$ one has
\begin{equation}\label{tayh}
	f=f_l(z-z_i)^l+f_{2}(z-z_i)^{l+1}+\cdots,~~l\geq1\,.
\end{equation}
with $f_l\neq0$. Since we have required $A_t(z_i)\neq0$, the smoothness of~\eqref{vectoreom} gives $n\geq l$. The leading terms in the right and left hands of~\eqref{vectoreom} are, respectively, given by
\begin{equation}\label{tayeq}
	n(n+l-1)f_l\rho_n(z-z_i)^{n+l-2},\quad  -\frac{q^2e^{\chi(z_i)}\rho_n A_t(z_i)^2}{f_l}(z-z_i)^{n-l}\,.
\end{equation}
If $l=1$, one should have
\begin{equation}\label{loeq1}
n^2 f_1^2=-q^2e^{\chi(z_i)} A_t(z_i)^2\,.
\end{equation}
This is impossible because the left hand side is positive, while the right hand side is negative. On the other hand, if $l>1$, one must demand
\begin{equation}\label{loeq2}
  \frac{q^2e^{\chi(z_i)}\rho_n A_t(z_i)^2}{f_l}=0\,.
\end{equation}
This is also impossible since $\rho_n\neq 0$ and $A_t(z_i)\neq 0$.

Therefore, a smooth horizon of the hairy black hole~\eqref{ansatz} must have $A_t=0$ on that horizon. For the generalized cases of~\eqref{actionnew} with  smooth functions $X$ and $V$ and of~\eqref{GB} with Gauss-Bonnet term, the corrections to~\eqref{vectoreom} only have terms higher than $(z-z_i)^{n-l}$, and thus do not change our result.

\providecommand{\href}[2]{#2}\begingroup\raggedright\endgroup

\end{document}